\documentclass{emulateapj}
\usepackage{apjfonts}
\usepackage{epsf}
\begin{document}
\slugcomment{}
\shortauthors{J. M. Miller et al.}
\shorttitle{GX 340$+$0}

\title{An Ultra-Fast X-ray Disk Wind in the Neutron Star Binary GX 340$+$0}

\author{J.~M.~Miller\altaffilmark{1},
  J.~Raymond\altaffilmark{2},
  E.~Cackett\altaffilmark{3},
  V.~Grinberg\altaffilmark{4},
  M.~Nowak\altaffilmark{4}
}
 
\altaffiltext{1}{Department of Astronomy, University of Michigan, 1085
  South University Avenue, Ann Arbor, MI 48109-1104, USA,
  jonmm@umich.edu}

\altaffiltext{2}{Harvard-Smithsonian Center for Astrophysics, 60
  Garden Street, Cambridge, MA 02138, USA}

\altaffiltext{3}{Department of Physics \& Astronomy, Wayne State
  University, 666 W. Hancock St, Detroit, MI 48201, USA}

\altaffiltext{4}{Massachusetts Institute of Technology, Kavli
  Institute for Astrophysics, Cambridge, MA 02139, USA}


\label{firstpage}

\begin{abstract}
We present a spectral analysis of a brief {\it Chandra}/HETG
observation of the neutron star low-mass X-ray binary GX~340$+$0.  The
high-resolution spectrum reveals evidence of ionized absorption in the
Fe K band.  The strongest feature, an absorption line at approximately
6.9~keV, is required at the 5$\sigma$ level of confidence via an
$F$-test.  Photoionization modeling with XSTAR grids suggests that the
line is the most prominent part of a disk wind with an apparent
outflow speed of $v = 0.04c$.  This interpretation is preferred at the
$4\sigma$ level over a scenario in which the line is H-like Fe XXVI at
a modest red-shift.  The wind may achieve this speed owing to its
relatively low ionization, enabling driving by radiation pressure on
lines; in this sense, the wind in GX 340$+$0 may be the stellar-mass
equivalent of the flows in broad absorption line quasars (BALQSOs).
If the gas has a unity volume filling factor, the mass ouflow rate in
the wind is over $10^{-5}~M_{\odot}~{\rm yr}^{-1}$, and the kinetic
power is nearly $10^{39}~{\rm erg}~{\rm s}^{-1}$ (or, 5--6 times the
radiative Eddington limit for a neutron star).  However, geometrical
considerations -- including a small volume filling factor and low
covering factor -- likely greatly reduce these values.
\end{abstract}

\section{Introduction}
Accretion onto neutron stars with low magnetic fields is expected to
be similar to accretion onto black holes, or at least black holes with
low spin parameters.  Time scales and line shifts should simply scale
with mass, and with the depth of the inner edge of the disk within the
potential well.  Observations appear to confirm this expectation:
quasi-periodic oscillations scale in the expected manner (e.g.,
Wijnands \& van der Klis 1999), and relativistic lines from neutron
stars place interesting limits on stellar radii but are not as extreme
as lines from spinning black holes (e.g. Cackett et al.\ 2008, 2010;
Di Salvo et al.\ 2009; Papitto et al.\ 2009; Miller et al.\ 2013).

Accretion flows onto neutron stars and black holes might be even more
similar, far from the compact object.  X-ray disk winds from
stellar-mass black holes are emerging as important facets of the
accretion flow; in some cases, the mass loss rate can rival or exceed
the accertion rate in the inner disk (e.g. King et al.\ 2012, 2015;
Miller et al.\ 2015).  Such winds appear to arise between
$10^{2-4}~{\rm GM/c^{2}}$ from the black hole, depending on
particulars.  Ionized X-ray absorption is commonly observed in neutron
star binaries, but evidence of outflowing gas has lagged the rapid
progress being made in stellar-mass black holes.  Nevertheless,
evidence of disk winds in neutron star X-ray binaries is growing, with
flows in GX 13$+$1 (e.g., Ueda et al.\ 2004) and IGR J17480$-$2446
(Miller et al.\ 2011) marking two prominent examples.

In some moderate-resolution spectra of active galactic nuclei (AGN)
extremely fast outflows with very high mass fluxes have been observed
(e.g., Tombesi et al.\ 2010).  These winds may contribute
significantly to feedback and the co-evolution of massive black holes
and host galaxies.  In stellar-mass black holes, outflow speeds of
0.01--0.05$c$ are detected at gratings resolution (e.g., King et
al.\ 2012, 2015, Miller et al.\ 2015).  Such flows have not previously
been reported in the steady emission of neutron star X-ray binaries
(but see Pinto et al.\ 2014 for evidence of fast outflows during
bursts).

GX 340$+$0 is a neutron star X-ray binary, located close to the
Galactic Plane.  It is known to be a low-mass X-ray binary and a ``Z''
source based on its behavior in X-ray color-color diagrams (Hasinger
\& van der Klis 1999).  Cackett et al.\ (2010) analyzed an {\it
  XMM-Newton} observation of GX~340$+$0, and found evidence of a
relativistic iron line from the inner disk, strongly impacted by X-ray
absorption consistent with H-like Fe XXVI (also see D'Ai et
al.\ 2009).  Lavagetto et al.\ (2004) also found X-ray absorption in a
{\it BeppoSAX} spectrum of GX~340$+$0; this was modeled with a
Gaussian at 6.8~keV.  This energy lies between Fe XXV (6.70~keV) and
Fe XXVI (6.97~keV), and would imply a large velocity shift.
In order to better understand the nature of the ionized X-ray
absorption in GX 340$+$0 at high resolution, we have analyzed archival
{\it Chandra}/HETG spectra of the source.

\section{Observations and Reduction}
{\it Chandra} has observed GX~340$+$0 on four occasions, but only
spectra from ObsID 1922 show evidence of a disk wind; it is the sole
focus of this paper.  Observation 1922 started on 2001-08-09 at
12:35:11 (UT).  After our reduction, we find that a net exposure time
of 5.8~ks was achieved.

The data were reduced using CIAO version 4.7 and the associated
calibration files.  We downloaded the full observation from the
{\it Chandra} archive.  The standard sequence of routines,
{\em tgdetect}, {\em tg\_create\_mask}, {\em tg\_resolve\_events}, and
{\em tgextract} were run to produce first-order spectral files.  When
executing {\em tg\_create\_mask}, we set the parameter
``width\_factor\_hetg'' to have a value of 17, rather than the default
of 35.  This reduces the width of the HEG extraction regions, with the
effect of better selecting true first-order HEG photons and enabling
the extraction of spectra that carry out to 10~keV.  The tools {\em
  mkgrmf} and {\em fullgarf} tasks were run to create response files.
Finally, {\em add\_grating\_spectra} was run to combine the
first-order spectra and responses, and the FTOOL ``grppha'' was used
to group the co-added first-order HEG spectrum to require at least 10
counts per bin.

\section{Analysis and Results}
The spectra were fit using XSPEC version 12.8.2 (Arnaud 1996).  The
``Churazov'' weighting scheme was applied in minimizing the $\chi^{2}$
fitting statistic.  All errors reported in this work reflect parameter
values at the $1\sigma$ confidence limits.

Cackett et al.\ (2010) report an interstellar column density of
$N_{\rm H, ISM} = 0.9-1.1 \times 10^{23}~{\rm cm}^{-2}$ along the line
of sight to GX~340$+$0.  D'Ai et al.\ (2009) report similarly high
values.  This greatly diminishes the low energy flux from this source.
Owing to the lack of sensitivity at low energy, and also owing to chip
gaps creating calibration uncertainties in the 3--4~keV range, our
analysis was restricted to the combined HEG spectrum and the 4--10~keV
band.

An initial fit to the spectrum with an absorbed disk blackbody model
($tbabs\times diskbb$; Wilms et al.\ 2000, Mitsuda et al.\ 1984)
captures the continuum fairly well, but leaves strong residuals in the
Fe K band.  The resultant fit statistic is $\chi^{2}/\nu = 686.37/566
= 1.213$.  The implied column density is $N_{\rm H} = 1.5\times
10^{23}~{\rm cm}^{-2}$, and the disk temperture is $kT = 0.85$~keV.

Figure 1 depicts this simple model for the spectrum, as well as the
data/model ratio, and deviations from the continuum as measured in
units of $\Delta\chi^{2}$.  There is a broad flux excess above the
continuum in the 5.5--6.5~keV range; this is likely an Fe K emission
blurred by effects in the inner accretion disk.  D'Ai et al.\ (2009)
and Cackett et al.\ (2010) both detected this disk line in independent
analyses.  The most striking and significant feature, however, is an
apparent absorption line at 6.9~keV.  The addition of a Gaussian model
for this line improves the overall fit to $\chi^{2}/\nu = 641.32/563 =
1.139$.

We performed 2000 Monte Carlo simulations of an absorbed disk spectrum
with no line features.  We fitted and conducted error bar searches on
the resulting spectra with a model including a Gaussian line (emission
or absorption) {\em anywhere} in the 4--10 keV range.  The detection
of a line feature with equal or greater strength is 99.6\% unlikely.
This represents a conservative probability for a blind search;
in contrast, our search focused on the Fe K band.  The change in
$\Delta \chi^{2}$ indicates a line significant at the $5.6\sigma$
level of confidence, as measured by an $F$-test.  Counting the number
of resolution elements in the 6.7--7.3~keV band as independent trials,
the significance is reduced to $5.1\sigma$.

The best-fit line energy is $6.94\pm 0.02$ keV. The rest-frame energy
of the He-like Fe XXV resonance line is 6.700~keV, and the rest-frame
energy of the H-like Fe XXVI resonance line is 6.970~keV (Verner et
al.\ 1996).  Thus, the line could represent absorption in a modest
inflow, or in a strong outflow.  There are weaker features in the
6.5--6.8~keV range and in the 7.0--7.5~keV range that may aid a
self-consistent determination of whether the gas is an inflow or
outflow.

To address this question, physical self-consistency is required;
line-by-line fitting is not sufficient.  We therefore constructed a
grid of XSTAR photoionized absorption models (e.g., Kallman \&
Bautista 2001).  A turbulent velocity of $300~{\rm km}~{\rm s}^{-1}$
and solar abundances were assumed.  A nominal gas density of $n =
10^{14}~{\rm cm}^{-3}$ was also assumed.  A
covering factor must also be specified; we selected $\Omega/4\pi =
0.5$ based on examples of similar absorption in stellar-mass black
holes (e.g., Miller et al.\ 2015).  Based on the disk blackbody
fit, an input spectrum with $kT = 0.9$~keV and a 2--30~keV
luminosity of $L = 4.6\times 10^{37}~{\rm erg}~{\rm s}^{-1}$ was
assumed (based on the unabsorbed flux and an assumed distance of
8.5~kpc; see Penninx et al.\ 1993) .  A grid of models spanning a
broad range in column density and ionization was generated
using the ``xstar2xspec'', and included in our subsequent
XSPEC fits as a multiplicative table model.  

When the grid is allowed to act on the blackbody continuum, it is a
$5\sigma$ improvement over a model without the absorption, giving
$\chi^{2}/\nu = 648.7/563 = 1.152$.  In this fit, $N_{\rm H, wind} =
3.3(7) \times 10^{22}~{\rm cm}^{-2}$ and ${\rm log}(\xi) = 3.1(1)$.
Importantly, a blue-shift of $v = -0.0395(5)c$ is measured.  A model
wherein the flow is required to have either zero shift or a red-shift,
is significantly worse: $\chi^{2}/\nu = 671.86/563 = 1.193$.  This is
true despite the fact that the ionization parameter moves up to ${\rm
  log}(\xi) = 5$, as expected if Fe XXVI must be very strong relative
to Fe XXV.  An F-test prefers the blue-shifted model over the
zero-shift model at the $4.5\sigma$ level of confidence.  This result
does not depend on the interstellar column density; the same
blue-shift results when a value of $N_{\rm H, ISM} = 4\times
10^{22}~{\rm cm}^{-2}$ is enforced (though the overall fit is not as
good).

This indicates that the absorption is best associated with
an rapid outflow.  However, it is possible that
the overall spectral model is still too simple.  Prior work has found
evidence of a relativistic disk line in GX~340$+$0, and the
broad flux excess in Figure 1 appears to confirm this feature in the
combined {\it Chandra}/HEG spectrum.  We there added a simple
relativistic line, ``diskline'' (Fabian et al.\ 1989), in
the next model.  ``Diskline'' is characterized in terms of a line
energy (restricted to the 6.40--6.97~keV range for Fe I--XXVI in our
fits), an disk emissivity index ($J \propto r^{q}$, bounded to lie in
the $-3 \leq q \leq -1$ range), an inner disk radius (measured in
units of ${\rm GM/c^{2}}$), an outer disk radius (fixed at $1000~{\rm
  GM/c^{2}}$), the inner disk inclination (bounded between $20^{\circ}
\leq \theta \leq 45^{\circ}$) and a flux normalization.

A model consisting of $tbabs\times xstar\_abs\times (diskline +
diskbb)$ improves the fit at the $5.6\sigma$ level of confidence, to
$\chi^{2}/\nu = 599.1/558 = 1.073$.  The absorption is not
substantially changed, a blue-shift of $v = -0.0395(6)c$ is again
measured ($N_{\rm H, wind} = 2.8\pm 0.1\times 10^{22}~{\rm cm}^{-2}$,
${\rm log}\xi = 3.1\pm 0.1$).  The measured diskline parameters
include ${\rm E} = 6.4^{+0.1}$~keV, $q = -3^{+0.1}$, $R_{in} =
6.0^{+0.3}$, $\theta = 20^{+1}$, and a normalization of $K = 0.010(1)$
(translating into an equivalent width of $W = 310\pm 30$~eV).  This
model is preferred over one including a disline but disallowing
blue-shifts at the $4\sigma$ level of confidence ($\chi^{2}/\nu =
618.0/558$).

Weak residuals remain in the Fe K band, suggesting that the wind is
more complex than our single-zone model.  A model with three zones
achieves only modest improvements in the fit statistic, giving
$\chi^{2}/\nu = 590.1/552 = 1.069$.  This fit accounts for weaker
residuals in the Fe K band, but also fits some apparent lines in the
4--5~keV band that can be associated with He-like Ca XIX and H-like Ca
XX (see Figures 2 and 3).  We regard this model as our best-fit model;
it is fully described in Table 1.  The best model wherein all three
components cannot be blue-shifted gives $\chi^{2}/\nu = 637.9/552$ (a
$6.6\sigma$ difference for a change of one free parameter); the best
three-component model wherein only one component is disallowed a
blue-shift gives $\chi^{2}/\nu = 621.4/552$ (a $5.3\sigma$ difference
for a change of one free parameter).

The data do not require re-emission from the wind; dynamical
broadening of such emission can potentially give radius constraints
(e.g. Miller et al.\ 2015, 2016).  Winds do not have to be launched
with the local escape speed; rather, they can be accelerated
continually, or once certain conditions obtain.  However, within this
framework, an outflow velocity of $v = 0.04c$ corresponds to a
launching radius of $r \simeq 1250~{\rm GM/c^{2}} \simeq 2.6\times
10^{8}$~cm (assuming a neutron star of $1.4~M_{\odot}$).

The mass outflow rate in each component can be calculated by starting
with the standard formula, $\dot{M} = 4\pi r^{2} \rho v$.  Adjusting
for a non-spherical flow, and writing in terms of number density, the
equation becomes $\dot{M} = \Omega \mu m_{p} n r^{2} v$ (where
$\Omega$ is the covering factor, $\mu = 1.23$ is the mean atomic
weight, $m_{p}$ is the mass of the proton, and $n$ is the number
density).  Using the ionization parameter (recall, $\xi = L/nr^{2}$),
it is possible to write the mass outflow rate without assuming a
density: $\dot{M} = \Omega \mu m_{p} (L/\xi) v$.  The kinetic power in
the outflow is then just $L_{kin} = 0.5 \dot{M} v^{2}$.

For the fast $v = 0.04c$ component detected in GX~340$+$0, the mass
outflow rate is $\dot{M} \simeq 1.1\times 10^{21}~{\rm g}~{\rm
  s}^{-1}$, or $\dot{M} \simeq 1.8\times 10^{-5}~M_{\odot}~{\rm
  yr}^{-1}$.  This is a very high mass flux; the implied inflow
rate at the inner disk is just $5\times 10^{17}~{\rm g}~{\rm s}^{-1}$,
assuming an efficiency of $\eta = 0.1$.  The implied power in the fast
component is even more extreme: $L_{kin} \simeq 8.1\times 10^{38}~{\rm
  erg}~{\rm s}^{-1}$.  This implies that the mechanical power exceeds
the radiative eddington limit by a factor of 5--6.

The estimates can be interpreted as the mass outflow rate and kinetic
power {\em divided by the filling factor}.  If the volume filling
factor is, e.g., $f\simeq 10^{-2}$ (commensurate with some AGN; see,
e.g., Blustin et al.\ 2005), the actual outflow rate is sub-Eddington
and the mass outflow rate is also reduced in direct proportion.
If the launching radius derived by associating the
observed wind speed with a local escape velocity is used, a very high
density value results ($n \simeq 1.1\times 10^{18}~{\rm cm}^{-3}$),
and $f = N_{\rm H, wind}/nr$ implies $f \simeq 1.6\times 10^{-4}$.  This would
reduce the mass outflow rate and kinetic power in GX 340$+$0 by four
orders of magnitude.  The mass outflow rate would then agree with the
inferred mass accretion rate at the inner disk, to within a factor of
a few.  However, as noted above, there is no requirement that the
observed wind speed is a local escape velocity.  If the filling factor
is not low, it may be the case that GX~340$+$0 was observed in a
super-Eddington phase.

\section{Discussion and Conclusions}
We have analyzed an archival {\it Chandra}/HETG spectrum of the
neutron star low-mass X-ray binary GX 340$+$0.  The spectra reveal
strong evidence of a single strong line close to 6.9~keV, and several
weaker aborption lines.  Self-consistent photoionization modeling
establishes that the feature is most likely produced in a complex disk
wind, with a component that is blue-shifted by $v = 0.04c$.  Even if
the wind has a low volume filling factor, its mass outflow rate and
kinetic power would still rank among the highest -- or as {\it the}
highest -- yet detected from a neutron star or stellar-mass black
hole.  In this section, we discuss the physical processes by which the
wind may be driven, and compare the outflow to extreme winds detected
in other sources.

Disk winds detected in {\it Chandra} observations of V404 Cyg and IGR
J17091$-$3634 may offer the best points of comparison for GX~340$+$0.
The 2015 outburst of V404 Cyg was particularly extreme, and it is
possible that the mass accretion rate was highly super-Eddington in
some phases.  In spectra of V404 Cyg, the observed velocity shifts
exceed $0.01c$ and the implied mass outflow rate is approximately
$\dot{M}_{wind} \simeq 10^{-5}~M_{\odot}~{\rm yr}^{-1}$ (assuming a
unity filling factor; King et al.\ 2015).  Spectra of IGR
J17091$-$3624 reveal a wind with two components, with speeds of
$0.03c$ and $0.05c$.  Here again, the total outflow rate would
approach $10^{-5}~M_{\odot}~{\rm yr}^{-1}$ for a unity filling factor,
but the filling factor may be of order $10^{-4}$ (King et al.\ 2012).

If the filling factor of the wind in GX~340$+$0 is not small, the
outflow power may actually exceed the observed radiative luminosity.
This would likely signal super-Eddington accretion in GX~340$+$0.  The
wind would then be driven by electron scattering pressure.  At least
one ultra-luminous X-ray (ULX) source is powered by a neutron star
(Bachetti et al.\ 2014), and population considerations suggest that a
number of ULXs may harbor neutron stars (King \& Lasota 2016).
However, a super-Eddinton flow is difficult to reconcile with the
detection of a relativistic disk line: the central engine should be
blocked by a super-Eddington photosphere.  Moreover, the observed
column density in the outflow is well below $N_{\rm H} \simeq
10^{24}~{\rm cm}^{-2}$ (see Table 1).

Unlike the disk winds that are typically observed in stellar-mass
black holes, the wind in GX~340$+$0 has components with an ionization
parameter below $\xi = 10^{3}~{\rm erg}~ {\rm cm}~ {\rm s}^{-1}$.
Simulations have identified this as a threshold below which
radiation pressure on lines can drive disk winds (e.g., Proga 2003).
Directly tapping into the radiative luminosity may help to explain why
the outflow in GX~340$+$0 has a high speed.  In this sense, the
observed wind may be similar to the extreme outflows in BALQSOs (e.g.,
Arav et al.\ 2001).

In BALQSOs, however, geometric shielding is required to
keep the gas from becoming over-ionized by X-rays.  It is unclear how
such shielding might be achieved in an X-ray binary, particularly when
the relativistic emission line in this system indicates a clear view
of the inner disk (e.g., D'Ai et al.\ 2009, Cackett et al.\ 2010).  It
may be the case that a combination of mechanisms, plausibly including
radiative pressure, thermal driving (e.g. Begelman et al.\ 1983), and
magnetic processes (e.g. Blandford \& Payne 1982; Proga 2003) act to
drive the wind in GX~340$+$0.  It is notable that the inclination
angle of $\theta = 35^{\circ}\pm 1^{\circ}$ indicated by the
relativistic line, is commensurate with the optimal angle for driving
magnetocentrifugal winds (Blandford \& Payne 1982).

Figure 4 shows the 1.5--12.0~keV {\it RXTE}/ASM light curve of
GX~340$+$0, spanning intervals near the observation in which we have
detected a strong wind, and two subsequent {\it Chandra} observations
that appear to lack strong absorption.  The wind is detected in the
observation with the lowest flux, though the level is only 10--20\%
below the observations lacking wind absorption.  Hardness ratios might
reveal more information, but the ASM ratios are insensitive owing to
the very high column density along the line of sight to GX~340$+$0.
Triggered observations based on MAXI light curves and hardness ratios
may be able to reveal links between the state of the disk and wind
production in GX~340$+$0.

\clearpage

\begin{figure}
  \hspace{0.25in}
  \includegraphics[scale=0.6,angle=-90]{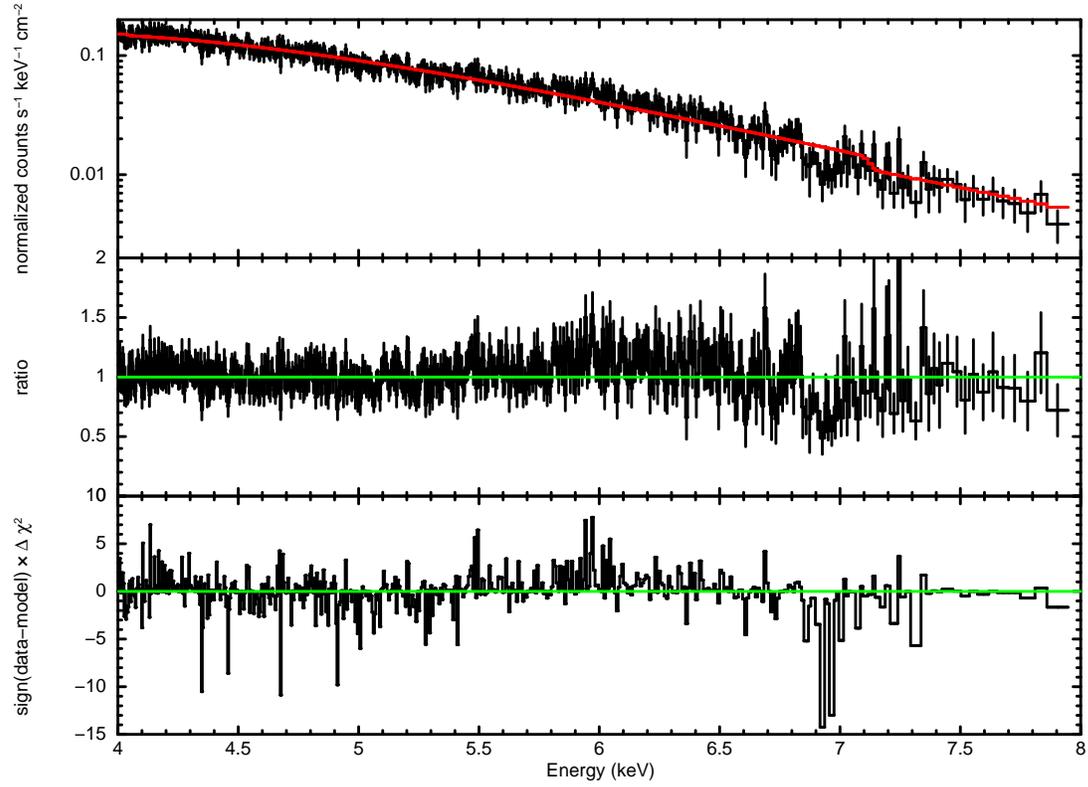}
\figcaption[t]{\footnotesize The combined first-order HEG spectrum of
  GX~340$+$0 from obsid 1922, fit with a simple disk blackbody
  continuum.  The top panel shows the spectrum and model (in red).
  The middle panel shows the data/model ratio.  The bottom panel shows
  the significance of departures from the continuum, in units of
  $\Delta\chi^{2}$.  The broad flux excess in the 5.5--6.5~keV range
  is likely a relativistically blurred Fe K emission line from the
  inner disk (see, e.g., D'Ai et al.\ 2009, Cackett et al.\ 2010).
  The absorption feature at 6.9 keV could potentially be H-like Fe
  XXVI (rest frame energy: 6.97~keV) at a modest red-shift, but it is
  more likely to be is either Fe XXV (rest frame energy: 6.70~keV) at
  a blue-shift of $\sim$0.04$c$.}
\vspace{0.25in}
\end{figure}
\medskip

\clearpage

\begin{table}[t]
\caption{Spectral Fitting Results}
\begin{footnotesize}
\begin{center}
\begin{tabular}{lllll}
\tableline
\tableline
Parameter & Zone 1 & Zone 2 & Zone 3  &  Continuum  \\
\tableline
${\rm N}_{\rm H, wind}~(10^{22}~{\rm cm}^{-2})$ & $4^{+3}_{-1}$ & $8^{+2}_{-1}$    & $4.7^{+0.8}_{-1.6}$    &  --  \\
log($\xi$)     &  $4.4^{+0.2}_{-0.6}$    &  $2.38(7)$     & $2.8^{+0.2}_{-0.1}$  &   --       \\
$v/c~(10^{-3})$     &  $-3.6^{+0.7}_{-1.1}$  & $-13.3^{+0.7}_{-0.4}$     &  $-39.7(5)$  &    --        \\
\tableline
$N_{\rm H,ISM}~ (10^{22}~{\rm cm}^{-2})$ &    --     &     --     &    --    &    $10.1^{+1.5}_{-0.3}$  \\
kT~(keV) &           --          &     --     &    --    &   0.86(1)  \\
diskbb~norm. &     --          &     --     &    --    &   $3000^{+700}_{-300}$ \\
${\rm E}_{diskline}$~(keV) & --          &     --     &    --    &  $6.40^{+0.03}$ \\
$q$ &     --          &     --     &    --    &   $-3.0^{+0.3}$ \\
$R_{in}~{\rm GM/c^{2}}$ --          &     --     &    --    &  -- &  $7.3^{+1.3}_{-0.7}$ \\
$\theta$           &     --     &    --    &  -- &  35(1) \\
${\rm Norm.}~ (10^{-2})$ --          &     --     &    --    &   -- &   $1.2^{+0.1}_{-0.3}$ \\
\tableline
\tableline
\end{tabular}
\vspace*{\baselineskip}~\\ \end{center} 
\tablecomments{The best-fit model to the HEG spectrum of GX~340$+$0 is
  detailed above.  The overall fit is good, though not formally
  acceptable: $\chi^{2}/\nu = 590.07/552 = 1.0690$.  Spectral
  continuum and wind properties are separated for clarity.  The
  continuum was described using a disk blackbody model.  A
  ``diskline'' component was included to describe a
  relativistic emission line.  The wind was modeled using three XSTAR
  components, characterized in terms of the wind column density,
  ionization parameter, and velocity shift ($N_{\rm H,wind}$, $\xi$,
  and $v/c$).  
}
\vspace{-1.0\baselineskip}
\end{footnotesize}
\end{table}
\medskip

\clearpage

\begin{figure}
  \hspace{0.25in}
  \includegraphics[scale=0.85]{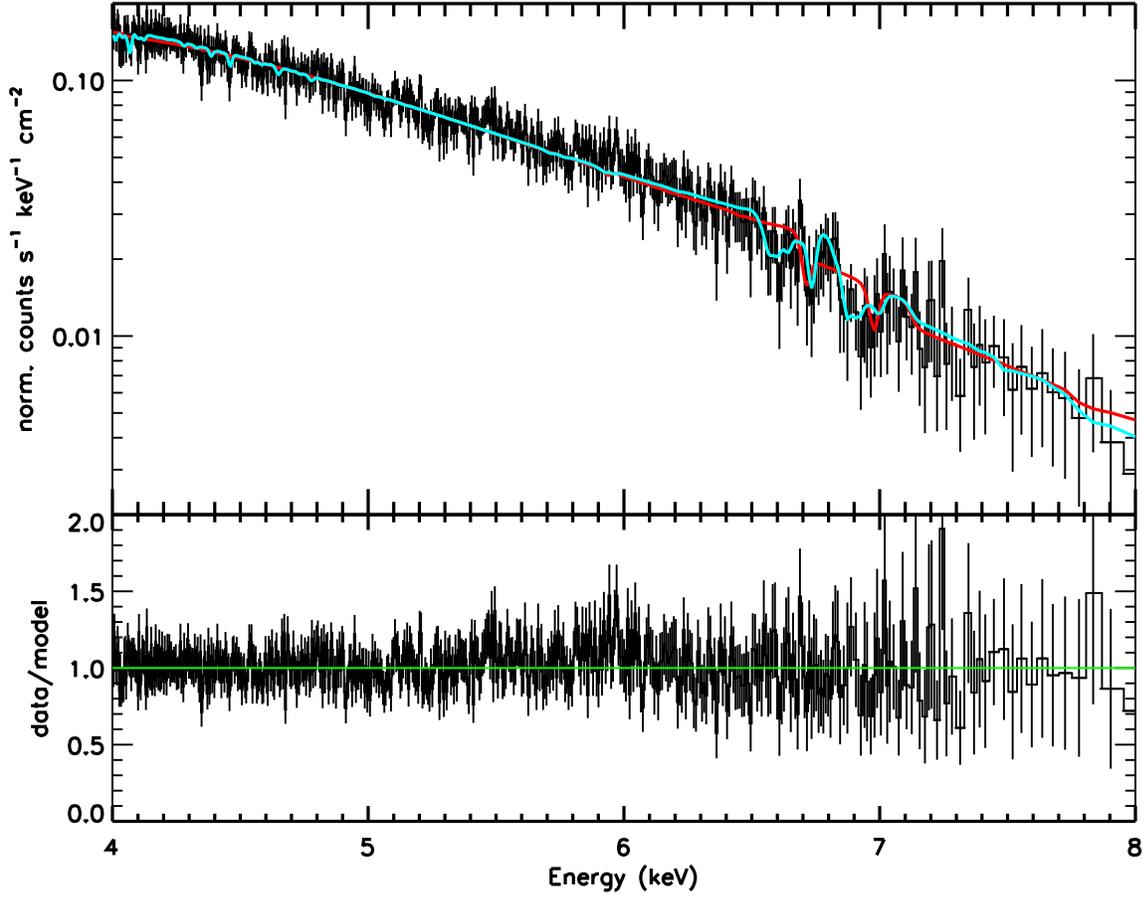}
\figcaption[t]{\footnotesize The combined first-order HEG spectrum of
  GX~340$+$0 from obsid 1922.  The best-fit model from Table 1, which
  includes strong blue-shifts, is shown in cyan.  For contrast, the
  best single-zone model that does not allow blue-shifts is shown in
  red.  The lower panel shows the ratio of the data to the best-fit model.}
\vspace{0.25in}
\end{figure}
\medskip

\begin{figure}
  \hspace{0.25in}
  \includegraphics[scale=0.85]{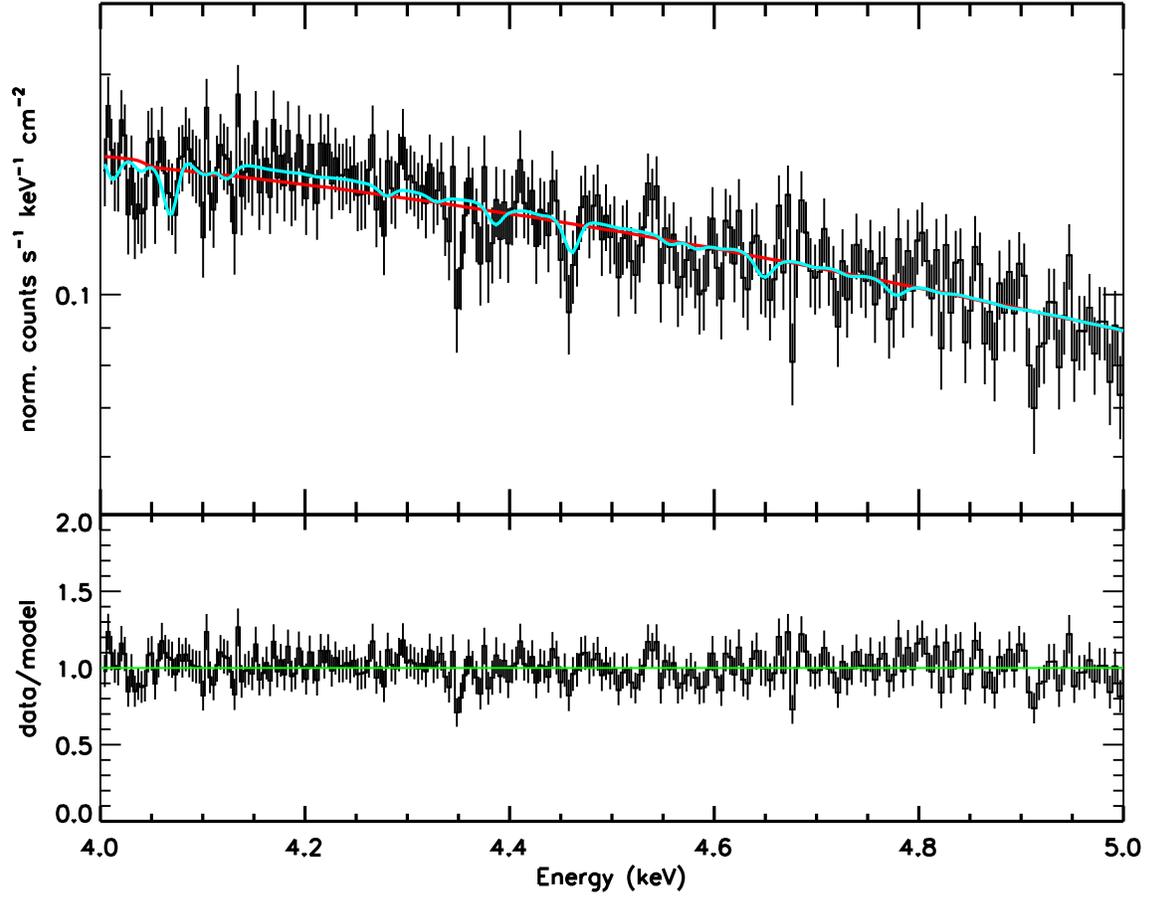}
\figcaption[t]{\footnotesize The combined first-order HEG spectrum of
  GX~340$+$0 from obsid 1922, with the best-fit model shown in cyan
  (see Table 1).  The model is able to explain some of the features in
  the 4--5~keV range in terms of blue-shifted absorption lines from
  He-like Ca XIX and H-like Ca XX.  For contrast, the model shown in
  red includes only one absorber, and blue-shifts were disallowed.
  The lower panel shows the ratio of the data to the best-fit model.}
\vspace{0.25in}
\end{figure}
\medskip

\begin{figure}
  \includegraphics[scale=0.9]{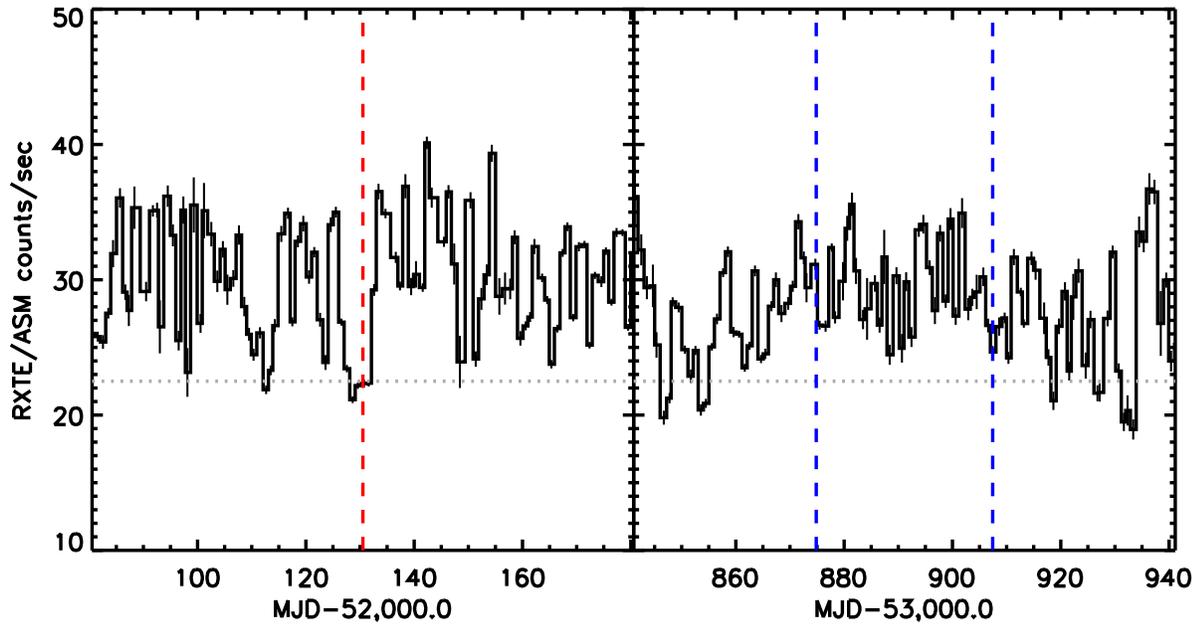}
  \figcaption[t]{\footnotesize Light curves of GX~340$+$0 from the
    RXTE/ASM (1.5--12.0~keV).  The left panel spans 100 days, centered
    on {\it Chandra} ObsID 1922 (dashed red line), in which the
    relativistic wind is detected.  The right panel spans 100 days
    including ObsIDs 6631 and 6632 (dashed blue lines), in which a
    wind is not apparent.  It is clear that ObsID 1922 was obtained at
    a lower source flux, and it is possible that such intervals should
    be targeted in the future to obtain a better understanding of the
    wind in GX~340$+$0.  Dashed, gray horizontal lines indicate the
    flux level at which ObsID 1922 was obtained.}
\vspace{0.25in}
\end{figure}
\medskip

\end{document}